\documentclass{article}
\usepackage{makecell}
\usepackage{spconf,amsmath,graphicx}
\usepackage{amsfonts}
\usepackage{multirow}
\usepackage{threeparttable}
\usepackage{booktabs}

\title{Exploring RWKV for Memory Efficient and Low Latency Streaming ASR}
\name{Keyu An, Shiliang Zhang}
\address{Speech Lab of DAMO Academy, Alibaba Group, China}
\begin{document}
\ninept
\maketitle
\begin{abstract}
Recently, self-attention-based transformers and conformers have been introduced as alternatives to RNNs for ASR acoustic modeling. Nevertheless, the full-sequence attention mechanism is non-streamable and computationally expensive, thus requiring modifications, such as chunking and caching, for efficient streaming ASR. In this paper, we propose to apply RWKV, a variant of linear attention transformer, to streaming ASR. RWKV combines the superior performance of transformers and the inference efficiency of RNNs, which is well-suited for streaming ASR scenarios where the budget for latency and memory is restricted. Experiments on varying scales (100h $\sim$ 10000h) demonstrate that RWKV-Transducer and  RWKV-Boundary-Aware-Transducer achieve comparable to or even better accuracy compared with chunk conformer transducer, with minimal latency and inference memory cost.
\end{abstract}
\section{Introduction}
\label{sec:intro}
\begin{figure}[!t]
\label{fig:comp}
  \centering
  \includegraphics[width=0.95\linewidth]{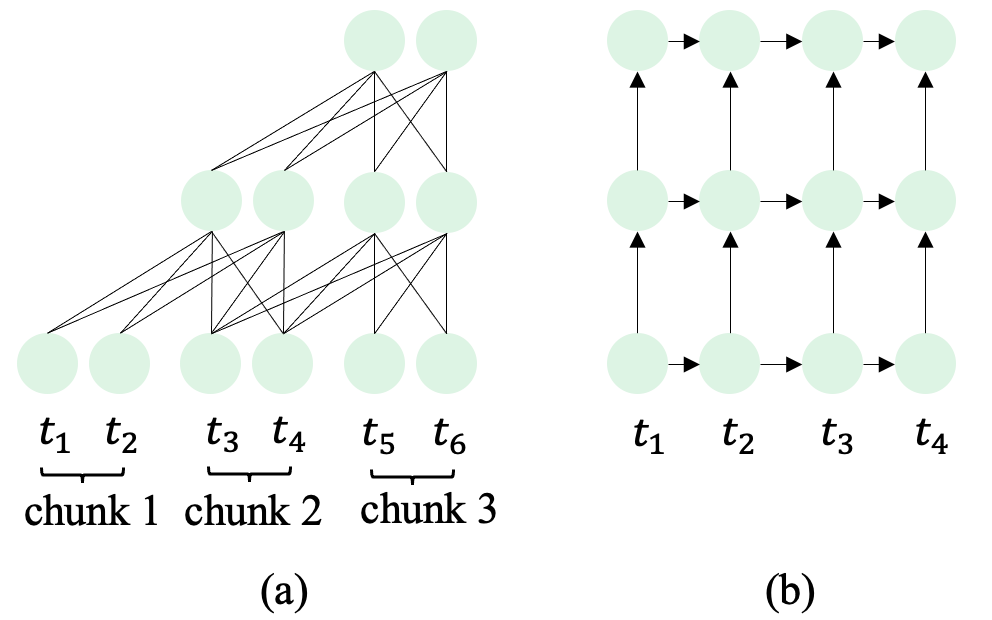}
  \caption{(a) chunk-based self-attention encoder. The current output is dependent on the input frame within the current chunk, and the representations of the previous chunks. (b) RWKV-based encoder. The current output only relies on the current input frame and the representations of the last frame.}
  \vspace{-0.25cm}
\end{figure}

Recently, self-attention-based neural networks such as Transformer~\cite{self-attention} and Conformer~\cite{Conformer} have been widely used for acoustic modeling in automatic speech recognition (ASR)~\cite{Conformer,transformer-transducer, Speech-Transformer} due to its superior performance compared to conventional RNN encoders~\cite{karita2019comparative}. 
Self-attention captures temporal dependencies in a sequence by computing pairwise attention scores between each input in a sequence, and thus is capable of leveraging contextual information for acoustic modeling, regardless of the sequence length. 
However, the full-sequence attention mechanism is inherently unsuitable for streaming ASR, where each word must be recognized shortly after it is spoken. 
To address it, causal self-attention~\cite{transformer-transducer, dual-mode-asr}, where the current frame only attends to the left context, and chunk-based self-attention~\cite{u2++, cuside}, where the current frame only attends to left context and a limited number of right context inside a chunk (Figure~\ref{fig:comp}(a)), are proposed to make the self-attention based encoder streamable. 
In these models, however, representations of the history input need to be stored in the cache to be reused as an extended context for the current output computation, which increases the memory consumption at inference, especially in applications where long contextual information is needed. 
In the chunk-based model, an additional issue is that it increases the recognition latency, as the calculation for the current output needs to wait for the future input.

In this paper, we propose to apply linear attention transformers for memory-efficient and low-latency streaming ASR. Linear attention mechanisms are recently introduced as alternatives to softmax self-attention~\cite{linear_transformer, rwkv}, especially to simplify the attention score computation. 
In linear attention transformers, the dot product between the query and the key in self-attention is replaced with linearized operations, which precludes the quadratic space-time computational complexity. 
Moreover, linear attention permits an iterative implementation and can be computed auto-regressively, and thus is capable of modeling long-range dependencies with minimal memory cost and is naturally suitable for streaming applications. 
Inspired by the success of linear attention transformers in multiple NLP tasks~\cite{linear_transformer, rwkv}, we propose to adopt Receptance Weighted Key Value (RWKV)~\cite{rwkv}, a variant of linear attention, to the streaming ASR, especially as the streamable RNN-T acoustic encoder. 
Extensive experiments prove that RWKV performs comparably with chunk self-attention encoder in accuracy, but with lower latency and is more memory-efficient in inference, due to its RNN-like formulation (Figure~\ref{fig:comp}(b)). This indicates that RWKV can be a promising alternative to the commonly used chunk-based streaming ASR acoustic encoders.

The paper is organized as follows. 
Section~\ref{sec:related} outlines related work. Section~\ref{sec:methods}  describes RWKV as the streamable neural transducer and boundary-aware transducer encoder. Experiments are shown in Section~\ref{sec:exp}. Section~\ref{limit} discusses the limitations of the method and future works. Section~\ref{conclu} gives the conclusion.

\section{Related works}
\label{sec:related}
Recurrent neural networks (RNNs)~\cite{lstm} are conventionally used as encoders for many sequence generation tasks. RNN has two appealing characteristics. First, it's naturally streamable as it does not require future context. Second, RNN represents the history of observations in a compressed state vector, hence the much lower memory cost at inference. Nevertheless, the performance of RNNs is inferior to that of transformers due to the well-known vanishing gradient problem and bottlenecks on the expressiveness~\cite{karita2019comparative,Capacity}.

On the other hand, there have been various attempts to make self-attention-based acoustic encoders streamable, such as causal self-attention and chunk self-attention. Causal self-attention~\cite{transformer-transducer} masks the attention score to the right of the current frame to produce output conditioned only on the previous state history. In chunk self-attention~\cite{u2++, An2023bat}, all frames within a chunk have access to one another and frames from a number of prior chunks. Typically, chunk-based attentions yield better performance due to its usage of future context, at the cost of higher latency. In both causal self-attention and chunk self-attention, the representations of the history input need to be cached for the current frame/chunk output calculation. Despite some previous work proposing to reduce memory consumption by compressing the past information~\cite{rae2019compressive, Shi2020EmformerEM}, the storage cost is still nonnegligible.

Linear attention transformers~\cite{linear_transformer, rwkv} are recently proposed to combine the strengths of RNNs and Transformers. In linear attention transformers, the traditional attention score computation, i.e. $QK^{T}$, is replaced with linearized operations, which results in better time and memory complexity as well as a causal model that can perform sequence generation auto-regressively in linear time. While Linear attention transformers have been successfully applied to sequence generation tasks such as image generation and phoneme recognition ~\cite{linear_transformer}, we for the first time explore its applications in streaming ASR, and give a thorough comparison of linear attention based RWKV and chunk conformer as streamable RNN Transducer (RNN-T) and Boundary-aware Transducer (BAT) encoders.

\section{Methods}
In this section, we introduce RWKV as a streamable transducer and boundary-aware transducer (BAT) encoder. While we choose transducer and boundary-aware transducer (BAT) because transducer-like models show superior performance and are well-suited to streaming decoding, the proposed encoder can be easily applied to other streaming ASR architectures such as CTC~\cite{ctc} and monotonic chunk-wise LAS~\cite{mocha}.
\label{sec:methods}
\subsection{Neural transducer}
Given the label sequences ${\bf y} = (y_1, y_2, ..., y_U) \in \mathcal{Y}$ and the input sequence ${\bf x}=(x_1, x_2, ..., x_T)$, RNN Transducer (RNN-T)~\cite{rnn-t} gives the label distribution conditioned on the input sequence and previous label history. In the training stage, RNN-T maximizes the log-probability
$$\mathcal{L} = -{\rm log} {\rm Pr}({\bf y} | {\bf x}) =  -{\rm log} \sum_{{\bf a}\in \mathcal{B}^{-1}(y)} {\rm Pr}(\bf{a}| {\bf x}) $$
where ${\bf a} = (a_1, a_2, ..., a_{T+U}) \in \mathcal{Y} \cup \{\phi\}$ is the blank label $\phi$ augmented alignment sequence, and the mapping $\mathcal{B}$ is defined by removing $\phi$ in the input sequence. 

${\rm Pr}(\bf{a} | {\bf x})$ is further factorized as
\begin{equation}
  \begin{aligned}
& {\rm Pr}({\bf a} | {\bf x}) =  \sum_{i=1}^{T+U} {\rm Pr}(a_i|h_{t_i}, g_{u_i})
 \nonumber 
  \end{aligned}
\end{equation}
where ${\bf h} = (h_1, h_2, ..., h_T) = {\rm Enc({\bf x})}$ is the high-level representation produced by the encoder, and $g_u$ is the prediction vector computed by the prediction network,
\begin{equation}
g_u = {\rm PredictNet}({\bf y}_{[0: u-1]})
 \nonumber 
\end{equation}
, with the convention $y_0 = \phi$. The probability ${\rm  Pr}(\cdot | h_t, g_u)$ is typically implemented as the output of the joint network:
\begin{equation}
\label{eq:pr}
{\rm Pr}(\cdot | h_t, g_u) = {\rm softmax}[{\bf W}^{out} {\rm tanh}({\bf W}^{enc} h_t + {\bf W}^{pred} g_u + b)]
 \nonumber 
\end{equation}
\subsection{Boundary-aware transducer}
One drawback of the standard neural transducer is that it requires large time and computation resources in training. Specifically, RNN-T evaluates the joint network for all possible (t, u) pairs, which results in a 4-D lattice of shape (N, T, U, V), where N is the batch size, T is the output length of the acoustic encoder, U is the output length of the prediction network, and V is the vocabulary size. To address it, Boundary-aware transducer (BAT)~\cite{An2023bat} proposed to select certain (t, u) pairs for evaluation based on the audio-text alignment, which is generated by a CIF~\cite{cif} module on-the-fly in training. Thus, the memory usage for RNN-T loss calculation is reduced to (N, T, R, V), where R is a pre-defined parameter that controls the ranges of the tokens that will be evaluated for every time step.

\subsection{RWKV as streamable transducer encoder}
\begin{figure}[!t]
  \centering
  \includegraphics[width=0.95\linewidth]{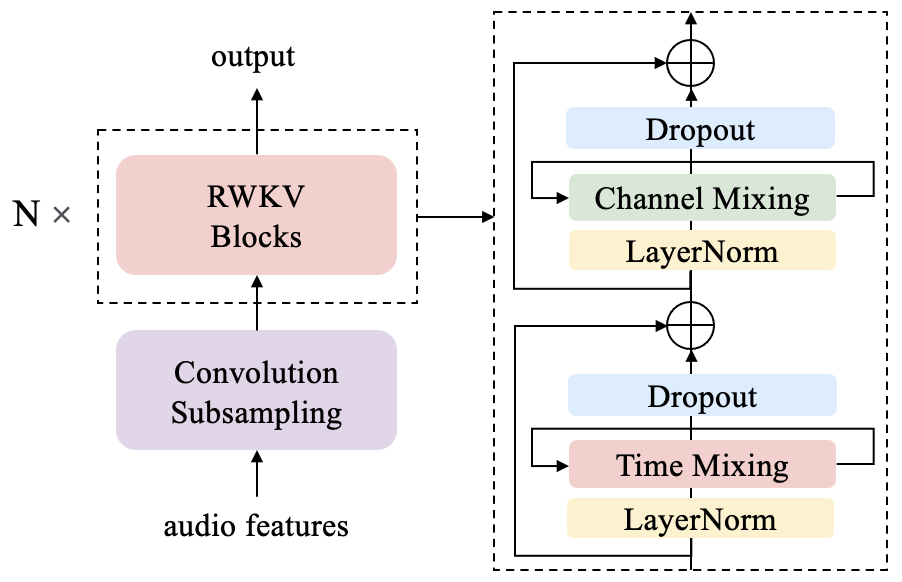}
  \caption{\textbf{RWKV as streamable ASR encoder.} A RWKV block comprises of a time mixing module and a channel mixing module with residual connections. A pre-layernorm layer and a post-dropout layer are adopted for each module.}
  \label{fig:rwkv}
  \vspace{-0.25cm}
\end{figure}
The RWKV encoder first processes the input with a convolution subsampling layer and then with a number of RWKV blocks. Each RWKV block is composed of a time-mixing and a channel-mixing sub-blocks with recurrent structures, as illustrated in Fig.~\ref{fig:rwkv}.
\subsubsection{time mixing module}
Given the input sequence ${\bf x}=(x_1, x_2, ..., x_T)$, where $T$ is the length of input features after convolution subsampling, the output of time mixing module ${\bf o}=(o_1, o_2, ..., o_T)$ is calculated as:
$$
o_t = W_o \cdot ({\sigma}(r_t)\odot  wkv_t) 
$$
where ${\sigma}(r_t)$ is the \textbf{receptance} vector at time step $t$, and $r_t$ is calculated as: 
$$
r_t = W_r \cdot ({\mu}_r x_t + (1 - {\mu}_r) x_{t-1})
$$
.$wkv_t$ plays the role of self-attention in transformers:
\begin{equation}
\label{wkv}
    wkv_t = \frac{\sum_{i=1}^{t-1} e^{-(t-1-i)w+k_i}v_i + e^{u+k_t}v_t }{\sum_{i=1}^{t-1} e^{-(t-1-i)w+k_i} + e^{u+k_t} }
\end{equation}
.$w$ is the channel-wise time decay vector for the previous input, and $u$ is the special weighting factor applied to the current input.
The \textbf{key} and \textbf{value} vectors are calculated as
$$
k_t = W_k \cdot ({\mu}_k x_t + (1 - {\mu}_k) x_{t-1})
$$
$$
v_t = W_v \cdot ({\mu}_v x_t + (1 - {\mu}_v) x_{t-1})
$$
$W_o \in \mathbb{R}^{d_{\rm io} \times d_{\rm att}}$ is the output projection matrix, where $d_{\rm io}$ is the input/output size, and $d_{\rm att}$ is the RWKV time mixing module size. $W_r \in \mathbb{R}^{d_{\rm att} \times d_{\rm io}}$, $W_k \in \mathbb{R}^{d_{\rm att} \times d_{\rm io}}$ and $W_v \in \mathbb{R}^{d_{\rm att} \times d_{\rm io}}$ are the projection matrix for the acceptance, key, and value respectively. ${\mu}_r$, ${\mu}_k$ and ${\mu}_v$ are time mix factors for the acceptance, key, and value respectively.

Note that $wkv_t$ is the weighted summation of the input in the interval $[1, t]$, which permits the causality in inference. Moreover Eq.~(\ref{wkv}) can be calculated recursively:
$$
wkv_t = \frac{a_{t-1} + e^{u+k_t}v_t}{b_{t-1} + e^{u+k_t}}
$$
where
$$
a_t = e^{-w}a_{t-1} + e^{k_t}v_t
$$
$$
b_t = e^{-w}b_{t-1} + e^{k_t}
$$
$$
a_0=b_0=0
$$
which enables efficient inference like RNNs.
\subsubsection{channel mixing module}
Given the input sequence ${\bf x^{\prime}}=(x_1^{\prime}, x_2^{\prime}, ..., x_T^{\prime})$, the output sequence of the the channel-mixing block is:
$$
o_t^{\prime} = {\sigma}(r_t^{\prime}) \cdot (W_v^{\prime}\odot  max(k_t^{\prime}, 0)^2) 
$$
where 
$$
r_t^{\prime} = W_r^{\prime} \cdot ({\mu}_r^{\prime} x_t^{\prime} + (1 - {\mu}_r^{\prime}) x_{t-1}^{\prime})
$$
$$
k_t^{\prime} = W_k^{\prime} \cdot ({\mu}_k^{\prime} x_t^{\prime} + (1 - {\mu}_k^{\prime}) x_{t-1}^{\prime})
$$
$W_r^{\prime} \in \mathbb{R}^{d_{\rm linear} \times d_{\rm io}}$ and $W_k^{\prime} \in \mathbb{R}^{d_{\rm linear} \times d_{\rm io}}$are the projection matrix for the acceptance and key respectively. $W_v^{\prime} \in \mathbb{R}^{d_{\rm io} \times d_{\rm linear}}$ is the channel-mixing matrix, and $d_{\rm linear}$ is the RWKV time mixing module size. ${\mu}_r^{\prime}$ and ${\mu}_k^{\prime}$ are time mix factors for the acceptance and key respectively. The channel mixing module is also causal as the calculation of $o_t^{\prime}$ only involves $x_t^{\prime}$ and $x_{t-1}^{\prime}$.
\subsubsection{RWKV block}
Given the input sequence ${\bf x}$, an RWKV block combines the time mixing module and channel mixing module using: 
$$
{\bf x}^{\prime} = {\bf x} + {\rm Dropout(TimeMixing(LayerNorm({\bf x})))}
$$
$$
{\bf x}^{\prime \prime} = {\bf x}^{\prime} + {\rm Dropout(ChannelMixing(LayerNorm({\bf x^{\prime}})))}
$$
Different from the original formulation~\cite{rwkv}, we add a Dropout layer before residual connection to avoid over-fitting.
\section{Experiments}
\label{sec:exp}
\begin{table}
\caption{The configurations of RWKV (small) and RWKV (large).}
\begin{threeparttable}
    \centering
    \scalebox{1}{
    \begin{tabular}{lcc}
    \toprule
    \textbf{Encoder} & \textbf{RWKV (S)} & \textbf{RWKV (L)} \\
    \midrule
    Input/output size $d_{\rm io}$& 512 & 640 \\
    Time Mixing Size $d_{\rm att}$ & 512 & 640 \\
    Channel Mixing Size $d_{\rm linear}$ & 2048 & 2560 \\
    Encoder Blocks $N$ & 18 & 18 \\
    Num Params (M) & 62 & 96 \\
    \bottomrule
    \end{tabular}}
\end{threeparttable}
\label{config}
\vspace{-0.25cm}
\end{table}

\begin{table*}
\caption{The latency, left context, and accuracy of different steaming models on AISHELL-1 (CER) and Librispeech (WER).}
\begin{threeparttable}
    \centering
    \scalebox{1}{
    \begin{tabular}{llccccc}
    \toprule
    \multirow{2}{*}{\textbf{model}} & \multirow{2}{*}{\textbf{encoder}} & \multicolumn{1}{c}{\textbf{latency}} & \multicolumn{1}{c}{\textbf{left context }} &\multicolumn{1}{c}{\textbf{AISHELL-1}}    & \multicolumn{2}{c}{\textbf{LibriSpeech}} \\
      & & (ms) & (\#frames) &test &  test clean & test other   \\
    \midrule
    CTC + Att rescoring & chunk conformer~\cite{u2++} & 640 + $\Delta$ &  all history & 5.05 & 3.80 & 10.38 \\   
    Transducer & chunk conformer~\cite{huang2020improving} & 400 & 40 & 6.15 & - & -\\
    Transducer & streaming transformer~\cite{transformer-transducer} & 0 & 10 & - & 4.2 & 11.3  \\
    Transducer & streaming transformer~\cite{transformer-transducer} & 0 & 2 & - & 4.5 & 14.5  \\
    Transducer & causal conformer~\cite{dual-mode-asr} & 0 & all history & - & 4.6 & 9.9  \\
    Transducer & causal conformer + distill~\cite{dual-mode-asr} & 0 & all history & - & 3.7 & 9.2  \\
    Transducer & conv augmented LSTM~\cite{conv-rnnt} & 0  & 1 & - & 5.11 & 13.82  \\
    \midrule
    Transducer & chunk conformer & 640 & 16 & \textbf{6.04} & \textbf{3.58} &  \textbf{9.27} \\
    Transducer & chunk conformer & 320 & 8 & 6.32 & 4.19 & 10.84 \\
    Transducer   &    RWKV(S)  & 0 & 1 & 6.11 & 3.83 & 9.63 \\
    BAT       &  RWKV(S) & 0 & 1 &  6.11 & 3.90 & 9.56 \\
    \bottomrule
    \end{tabular}}
\end{threeparttable}
\label{rwkv_results}
\vspace{-0.25cm}
\end{table*}

\begin{table*}
\caption{The latency, left context, and accuracy of different steaming models on WenetSpeech (CER) and Gigaspeech (WER).}
\begin{threeparttable}
    \centering
    \scalebox{1}{
    \begin{tabular}{llcccccc}
    \toprule
    \multirow{2}{*}{\textbf{model}} & \multirow{2}{*}{\textbf{encoder}} & \multicolumn{1}{c}{\textbf{latency}} & \multicolumn{1}{c}{\textbf{left context}} & \multicolumn{3}{c}{\textbf{WenetSpeech}} & \multicolumn{1}{c}{\textbf{GigaSpeech}}\\
      &  & (ms) & (\#frames) & Dev &  Test\_Net & Test\_Meeting & test  \\
    \midrule 
    CTC + Att rescoring & chunk conformer~\cite{u2++} & 480 + $\Delta$ &  all history & - & - & -& 12.5\\ 
    CTC + Att rescoring & chunk conformer~\cite{u2++} & 640 + $\Delta$ & all history &  8.87 & 10.22 & 18.11 & -\\ 
    \midrule
    Transducer & chunk conformer & 640 & 16 & \textbf{9.42} & 12.33 & \textbf{18.96} & 13.06 \\
    Transducer & chunk conformer & 320 & 24 & 13.79 & 16.55 & 28.06 & \textbf{12.43} \\
    Transducer & chunk conformer & 320 & 8 & 14.84 & 17.83 & 31.10 &  13.06 \\
    Transducer   &    RWKV(S)  & 0 & 1 & 10.45 & 11.88 & 21.06 & 13.12\\
    BAT       &  RWKV(S)  & 0 &1 & 10.75 & 12.27 & 21.98 & 13.19 \\
    Transducer   &    RWKV(L)  & 0  & 1& 10.49 & \textbf{11.46} & 19.43 & -\\
    BAT       &  RWKV(L)  & 0 & 1 & 10.52 & 11.76 & 20.36 & -\\
    \bottomrule
    \end{tabular}}
\end{threeparttable}
\label{rwkv_results2}
\vspace{-0.25cm}
\end{table*}
\subsection{Experiment settings}
We conduct experiments on the openly available 170-hour Mandarin AISHELL-1~\cite{aishell1}, 960-hour English LibriSpeech~\cite{libri}, 10000-hour Mandarin WenetSpeech~\cite{wenetspeech} and 10000-hour English GigaSpeech~\cite{gigaspeech} datasets. The code will be available in FunASR~\cite{gao2023funasr}~\footnote{https://github.com/alibaba-damo-academy/FunASR}. 

For all datasets, we use 80-dim filterbanks as input. The input features are extracted on a window of 25ms with a 10ms shift, and then subsampled by a factor of 4 using the convolution subsampling layer. The configuration of the RWKV encoder is shown in the Table~\ref{config}. For comparison, we report the results of a chunk-attention-based conformer transducer. The conformer encoder has 12 layers. The convolution kernel size of the conformer is 15 and the number of attention heads, attention dimension, and feed-forward dimension are 8, 512, and 2048 respectively. The attention chunk size is 16 (i.e. 640ms) or 8 (i.e. 320ms). The total parameters are 90M. For boundary-aware transducers, the number of tokens that will be evaluated for every time step is set to 5. We pre-train the CIF module for several epochs so it can produce more accurate audio-text alignment at the early stage.

\subsection{Metrics}
In addition to the accuracy measured by word error rate (WER, for English tasks) and character error rate (CER, for Mandarin tasks), we also report latency and the left context the model requires to demonstrate the inference efficiency of different models, which are detailed below.

\textbf{Latency.} The latency is defined by the future context the model accesses. For the chunk-based model, the latency is defined as the time duration of the chunk, i.e. chunk size $\times$ frame subsampling factor $\times$ time per frame. For the models based on RNN, causal self-attention, and linear attention, the latency is 0 as their prediction does not depend on the future context.

\textbf{Left context.} The left context is the number of left frames the model accesses for the current output.  The left context is directly related to the memory consumption at inference, as the feature frame representations for the left context need to be cached for reuse. For the model that uses all history frames as left context, the memory and computation cost per timestep scales with the square of the current sequence length because attention must be computed for all previous timesteps. For the model based on RNN and linear attention, the left context is 1 as the output of the timestep $t$ is only dependent on the current input and the output at timestep $t-1$. For the model based on chunk self-attention, the left context is defined by the number of left frames, which is typically much larger than 1.

\subsection{Results}
The results on 170-hour AISHELL-1 and 960-hour LibriSpeech are shown in Table~\ref{rwkv_results}, and the results on 10000-hour WenetSpeech and 10000-hour GigaSpeech are shown in Table~\ref{rwkv_results2}. Results from related literature are listed for comparison. We also report the results for chunk conformers with different latency and left context configurations. It can be seen that 

1) For the self-attention-based model (transformer and conformer), a number of left contexts (10 frames $\sim$ infinite) is indispensable, and lack of left context would lead to a significant degradation in recognition accuracy~\cite{transformer-transducer}. Moreover, a trade-off between latency and CER/WER is observed as models with larger chunk sizes tend to have lower CER/WER.

2) RWKV-based transducer and BAT achieve close performance with chunk conformer Transducers, with much lower latency and inference memory cost. Notably, When the chunk-based models adopt a relatively smaller chunk size and limited left context, the RWKV-based transducer and BAT show significant accuracy superiority on the LibriSpeech and Wenetspeech datasets, which indicates that RWKV encoder can exploit the context information efficiently and is well-suited for scenarios where the budget for latency and memory is highly restricted. On LibriSpeech and GigaSpeech datasets, RWKV-based Transducer and BAT show comparable or even better results with the two-pass CTC + Attention model, where the final results are selected from the streaming CTC hypothesis reranked by a non-streaming full-context model.

3) While transducer and BAT perform comparably in most benchmarks, transducer achieves much better accuracy on the WenetSpeech meeting test, presumably because it's more difficult for BAT to locate the word boundary in the meeting environment. The benefit of BAT is that it reduces about 40\% overall training memory cost and about 25\% overall training time cost in our experiments.

4) While the LSTM-based streaming transducer has similar advantages on latency and memory cost~\cite{conv-rnnt}, the performances are much worse than the chunk conformer and RWKV-based models, which reveals the superiority of linear attention transformers in modeling long-term dependencies and is consistent with the findings in other sequence generation tasks~\cite{linear_transformer}.

\section{Limitations and future work}
\label{limit}
In our experiments, the accuracy of RWKV encoder outperforms chunk conformer with limited context, but is inferior to the chunk conformer with a large chunk size and unlimited left context. A possible direction for improvements is to enhance the RWKV encoder with more context information, e.g.,  add convolutions to capture local context for speech, or combine the RWKV encoder and chunk-based model to allow efficient modeling of the left context and access to the limited right context at the same time. 
\section{Conclusions}
\label{conclu}
In this paper, we apply RWKV, a variant of linear attention transformer, to streaming ASR. Compared to the causal conformer and chunk conformer, RWKV has a lower memory cost at inference as the history context needed to be cached is minimal. Moreover, the RWKV encoder has minimal latency as it does not require any future context. Extensive experiments on various languages and scales demonstrate that RWKV-Transducer and RWKV-Boundary-Aware-Transducer achieve comparable or better performances with chunk conformers in accuracy, and serve well for scenarios where the budget for latency and memory is highly restricted.
\bibliographystyle{IEEEbib}
\bibliography{strings,refs}

\begin{thebibliography}{10}

\bibitem{self-attention}
Ashish Vaswani, Noam~M. Shazeer, Niki Parmar, Jakob Uszkoreit, Llion Jones,
  Aidan~N. Gomez, Lukasz Kaiser, and Illia Polosukhin,
\newblock ``Attention is all you need,''
\newblock in {\em Proc. Advances in Neural Information Processing Systems},
  2017, pp. 5998--6008.

\bibitem{Conformer}
Anmol Gulati, Chung-Cheng Chiu, James Qin, Jiahui Yu, Niki Parmar, Ruoming
  Pang, Shibo Wang, Wei Han, Yonghui Wu, Yu~Zhang, and Zhengdong Zhang,
\newblock ``Conformer: Convolution-augmented transformer for speech
  recognition,''
\newblock in {\em Proc. INTERSPEECH}, 2020, pp. 5036--5040.

\bibitem{transformer-transducer}
Qian Zhang, Han Lu, Hasim Sak, Anshuman Tripathi, Erik McDermott, Stephen Koo,
  and Shankar Kumar,
\newblock ``Transformer transducer: A streamable speech recognition model with
  transformer encoders and rnn-t loss,''
\newblock in {\em Proc. ICASSP}, 2020, pp. 7829--7833.

\bibitem{Speech-Transformer}
Linhao Dong, Shuang Xu, and Bo~Xu,
\newblock ``Speech-transformer: A no-recurrence sequence-to-sequence model for
  speech recognition,''
\newblock in {\em Proc. ICASSP}, 2018, pp. 5884--5888.

\bibitem{karita2019comparative}
Shigeki Karita, Nanxin Chen, Tomoki Hayashi, Takaaki Hori, Hirofumi Inaguma,
  Ziyan Jiang, Masao Someki, Nelson Enrique~Yalta Soplin, Ryuichi Yamamoto,
  Xiaofei Wang, et~al.,
\newblock ``A comparative study on transformer vs rnn in speech applications,''
\newblock in {\em Proc. ASRU}. IEEE, 2019, pp. 449--456.

\bibitem{dual-mode-asr}
Jiahui Yu, Wei Han, Anmol Gulati, Chung-Cheng Chiu, Bo~Li, Tara~N Sainath,
  Yonghui Wu, and Ruoming Pang,
\newblock ``Dual-mode {ASR}: Unify and improve streaming {ASR} with
  full-context modeling,''
\newblock in {\em Proc. ICLR}, 2021.

\bibitem{u2++}
Di~Wu, Binbin Zhang, Chao Yang, Zhendong Peng, Wenjing Xia, Xiaoyu Chen, and
  Xin Lei,
\newblock ``U2++: Unified two-pass bidirectional end-to-end model for speech
  recognition,''
\newblock {\em arXiv preprint arXiv:2106.05642}, 2021.

\bibitem{cuside}
Keyu An, Huahuan Zheng, Zhijian Ou, Hongyu Xiang, Ke~Ding, and Guanglu Wan,
\newblock ``{CUSIDE: Chunking, Simulating Future Context and Decoding for
  Streaming ASR},''
\newblock in {\em Proc. INTERSPEECH}, 2022, pp. 2103--2107.

\bibitem{linear_transformer}
Angelos Katharopoulos, Apoorv Vyas, Nikolaos Pappas, and Fran{\c{c}}ois
  Fleuret,
\newblock ``Transformers are {RNN}s: Fast autoregressive transformers with
  linear attention,''
\newblock in {\em Proc. ICML}, 2020, pp. 5156--5165.

\bibitem{rwkv}
Bo~Peng, Eric Alcaide, Quentin~G. Anthony, Alon Albalak, Samuel Arcadinho,
  Huanqi Cao, Xin Cheng, Michael Chung, Matteo Grella, G~Kranthikiran, Xuming
  He, Haowen Hou, Przemyslaw Kazienko, Jan Kocoń, Jiaming Kong, Bartlomiej
  Koptyra, Hayden Lau, Krishna Sri~Ipsit Mantri, Ferdinand Mom, Atsushi Saito,
  Xiangru Tang, Bolun Wang, Johan~Sokrates Wind, Stansilaw Wozniak, Ruichong
  Zhang, Zhenyuan Zhang, Qihang Zhao, Peng Zhou, Jian Zhu, and Rui Zhu,
\newblock ``Rwkv: Reinventing rnns for the transformer era,''
\newblock {\em arXiv preprint arXiv:2305.13048}, 2023.

\bibitem{lstm}
Sepp Hochreiter and J{\"u}rgen Schmidhuber,
\newblock ``Long short-term memory,''
\newblock {\em Neural computation}, vol. 9, no. 8, pp. 1735--1780, 1997.

\bibitem{Capacity}
Jasmine Collins, Jascha~Narain Sohl-Dickstein, and David Sussillo,
\newblock ``Capacity and trainability in recurrent neural networks,''
\newblock {\em arXiv preprint arXiv:1611.09913}, 2016.

\bibitem{An2023bat}
Keyu An, Xian Shi, and Shiliang Zhang,
\newblock ``Bat: Boundary aware transducer for memory-efficient and low-latency
  asr,''
\newblock in {\em Proc. INTERSPEECH}, 2023.

\bibitem{rae2019compressive}
Jack~W Rae, Anna Potapenko, Siddhant~M Jayakumar, and Timothy~P Lillicrap,
\newblock ``Compressive transformers for long-range sequence modelling,''
\newblock {\em arXiv preprint arXiv:1911.05507}, 2019.

\bibitem{Shi2020EmformerEM}
Yangyang Shi, Yongqiang Wang, Chunyang Wu, Ching feng Yeh, Julian Chan, Frank
  Zhang, Duc Le, and Michael~L. Seltzer,
\newblock ``Emformer: Efficient memory transformer based acoustic model for low
  latency streaming speech recognition,''
\newblock in {\em Proc. ICASSP}, 2021, pp. 6783--6787.

\bibitem{ctc}
Alex Graves, Santiago Fern\'{a}ndez, Faustino Gomez, and J\"{u}rgen
  Schmidhuber,
\newblock ``Connectionist temporal classification: Labelling unsegmented
  sequence data with recurrent neural networks,''
\newblock in {\em Proc. ICML}, 2006, p. 369–376.

\bibitem{mocha}
Chung-Cheng Chiu and Colin Raffel,
\newblock ``Monotonic chunkwise attention,''
\newblock in {\em Proc. ICLR}, 2018.

\bibitem{rnn-t}
A.~Graves,
\newblock ``Sequence transduction with recurrent neural networks,''
\newblock {\em arXiv preprint arXiv:1211.3711}, 2012.

\bibitem{cif}
Linhao Dong and Bo~Xu,
\newblock ``{CIF}: Continuous integrate-and-fire for end-to-end speech
  recognition,''
\newblock in {\em Proc. ICASSP}, 2020, pp. 6079--6083.

\bibitem{huang2020improving}
Mingkun Huang, Jun Zhang, Meng Cai, Yang Zhang, Jiali Yao, Yongbin You, Yi~He,
  and Zejun Ma,
\newblock ``Improving 1611.09913rnn transducer with normalized jointer
  network,''
\newblock {\em arXiv preprint arXiv:2011.01576}, 2020.

\bibitem{conv-rnnt}
Martin Radfar, Rohit Barnwal, Rupak~Vignesh Swaminathan, Feng-Ju Chang,
  Grant~P. Strimel, Nathan Susanj, and Athanasios Mouchtaris,
\newblock ``{ConvRNN-T: Convolutional Augmented Recurrent Neural Network
  Transducers for Streaming Speech Recognition},''
\newblock in {\em Proc. INTERSPEECH}, 2022, pp. 4431--4435.

\bibitem{aishell1}
Hui Bu, Jiayu Du, Xingyu Na, Bengu Wu, and Hao Zheng,
\newblock ``{AISHELL-1}: An open-source mandarin speech corpus and a speech
  recognition baseline,''
\newblock in {\em Proc. O-COCOSDA}, 2017, pp. 1--5.

\bibitem{libri}
Vassil Panayotov, Guoguo Chen, Daniel Povey, and Sanjeev Khudanpur,
\newblock ``Librispeech: An asr corpus based on public domain audio books,''
\newblock in {\em Proc. ICASSP}, 2015, pp. 5206--5210.

\bibitem{wenetspeech}
Binbin Zhang, Hang Lv, Pengcheng Guo, Qijie Shao, Chao Yang, Lei Xie, Xin Xu,
  Hui Bu, Xiaoyu Chen, Chenchen Zeng, et~al.,
\newblock ``Wenetspeech: A 10000+ hours multi-domain mandarin corpus for speech
  recognition,''
\newblock in {\em Proc. ICASSP}, 2022, pp. 6182--6186.

\bibitem{gigaspeech}
Guanbo~Wang Guoguo~Chen, Shuzhou~Chai and et~al.,
\newblock ``Gigaspeech: An evolving, multi-domain asr corpus with 10,000 hours
  of transcribed audio,''
\newblock in {\em Proc. Interspeech}, 2021.

\bibitem{gao2023funasr}
Zhifu Gao, Zerui Li, Jiaming Wang, Haoneng Luo, Xian Shi, Mengzhe Chen, Yabin
  Li, Lingyun Zuo, Zhihao Du, Zhangyu Xiao, and Shiliang Zhang,
\newblock ``Funasr: A fundamental end-to-end speech recognition toolkit,''
\newblock in {\em Proc. INTERSPEECH}, 2023.

\end{thebibliography}

\end{document}